\documentclass{article}
\usepackage{ijcai23}

\usepackage{times}
\usepackage{soul}
\usepackage{url}
\usepackage[hidelinks]{hyperref}
\usepackage[utf8]{inputenc}
\usepackage[small]{caption}
\usepackage{graphicx}
\usepackage{amsmath}
\usepackage{amsthm}
\usepackage{booktabs}
\usepackage{algorithm}
\usepackage{algorithmic}
\usepackage[switch]{lineno}

\usepackage{amsmath,graphicx,setspace,algorithm,algorithmic,multirow,cases,amssymb,subfigure}
\usepackage{color}
\usepackage{threeparttable}

\title{TEC-Net: Vision Transformer Embrace Convolutional Neural Networks for Medical Image Segmentation}

\author{
   Tao Lei, Rui Sun, Weichuan Zhang, Yong Wan, Yong Xia, and Asoke K. Nandi
    \affiliations
    Affiliation
    \emails
    email@example.com
}

\author{
Rui Sun$^1$
\and
Tao Lei$^{1,2}$
\and
Weichuan Zhang$^3$
\and
Yong Wan$^2$
\and
Yong Xia$^4$
\And
Asoke K. Nandi$^5$
\affiliations
$^1$Shaanxi Joint Laboratory of Artificial Intelligence, Shaanxi University of Science and Technology\\
$^2$Department of Geriatric Surgery, First Affiliated Hospital, Xi'an Jiaotong University\\
$^3$Griffith University, QLD, Australia\\
$^4$School of Computing, Northwestern Polytechnical University
$^5$Department of Electronic and Electrical Engineering, Brunel University London
\emails
leitao@sust.edu.cn,
siri0920@163.com,
Weichuan.Zhang@data61.csiro.au
docwanyong@xjtu.edu.cn
yxia@nwpu.edu.cn
Asoke.Nandi@brunel.ac.uk
}

\begin{document}

\maketitle

\begin{abstract}
The hybrid architecture of convolution neural networks (CNN) and Transformer has been the most popular method for medical image segmentation. However, the existing networks based on the hybrid architecture suffer from two problems. First, although the CNN branch can capture image local features by using convolution operation, the vanilla convolution is unable to achieve adaptive extraction of image features. Second, although the Transformer branch can model the global information of images, the conventional self-attention only focuses on the spatial self-attention of images and ignores the channel and cross-dimensional self-attention leading to low segmentation accuracy for medical images with complex backgrounds. To solve these problems, we propose vision Transformer embrace convolutional neural networks for medical image segmentation (TEC-Net). Our network has two advantages. First, dynamic deformable convolution (DDConv) is designed in the CNN branch, which not only overcomes the difficulty of adaptive feature extraction using fixed-size convolution kernels, but also solves the defect that different inputs share the same convolution kernel parameters, effectively improving the feature expression ability of CNN branch. Second, in the Transformer branch, a (shifted)-window adaptive complementary attention module ((S)W-ACAM) and compact convolutional projection are designed to enable the network to fully learn the cross-dimensional long-range dependency of medical images with few parameters and calculations. Experimental results show that the proposed TEC-Net provides better medical image segmentation results than SOTA methods including CNN and Transformer networks. In addition, our TEC-Net requires fewer parameters and computational costs and does not rely on pre-training. The code is publicly available at \textcolor{blue}{\itshape https://github.com/SR0920/TEC-Net.}
\end{abstract}

\section{Introduction}

image segmentation refers to dividing a medical image into several specific regions with unique properties. Medical image segmentation results can not only achieve abnormal detection of human body regions but also be used to guide clinicians. Therefore, accurate medical image segmentation has become a key component of computer-aided diagnosis and treatment, patient condition analysis, image-guided surgery, tissue and organ reconstruction, and treatment planning. Compared with common RGB images, medical images usually suffer from the problems such as high-density noise, low contrast and blurred edges. So how to quickly and accurately segment specific human organs and lesions from medical images has always been a huge challenge in the field of smart medicine.

The early traditional medical image segmentation algorithms are based on manual features designed by medical experts using professional knowledge~\cite{suetens2017fundamentals}. These methods have a strong mathematical basis and theoretical support, but these algorithms have poor generalization abilities for different organs or lesions of human body. Later, inspired by the full convolutional networks (FCN)~\cite{long2015fully} and the idea of encoder-decoder, Ronnebreger et al. designed U-Net~\cite{ronneberger2015u} network that was first applied to medical image segmentation. After that the U-shaped encoder-decoder structure receives widespread attention. At the same time, due to the small number of parameters and the good segmentation effect of U-Net, deep learning has made a breakthrough in medical image segmentation. Then a series of improved medical image segmentation networks are reported, such as 2D U-Net++~\cite{zhou2018unet++}, ResDO-UNet~\cite{liu2023resdo}, SGU-Net~\cite{lei2023sgu}, 2.5D RIU-Net~\cite{lv20222}, 3D Unet~\cite{cciccek20163d} and V-Net~\cite{milletari2016v}. The rapid development of CNN in the field of medical image segmentation is largely due to the scale invariance and inductive bias of convolution operation. Although this fixed receptive field improves the computational efficiency of CNN, it limits the ability of CNN to capture the long-range dependency relationship.

Aiming at the shortcomings of CNN in obtaining global features of images, Vaswani et al.~\cite{vaswani2017attention} proposed the vision Transformer architecture for image classification. The Transformer achieves a global representation of image information through complex spatial transformations and long-range dependency modeling, effectively solving the problem of CNN being only able to obtain local features of images. Currently, many methods based on Transformer have been applied to medical image segmentation, representative methods such as Swin-Unet~\cite{cao2023swin}, BAT~\cite{wang2021boundary}, Swin UNETR~\cite{tang2022self}, and UCTransNet~\cite{wang2022uctransnet}. These methods can be roughly divided into the pure Transformer architecture and the hybrid architecture of CNN and Transformer. The pure Transformer architecture realizes the long-range dependency modeling using self-attention. However, due to the lack of inductive bias of the Transformer itself, the traditional Transformer cannot be widely used on small-scale datasets like medical images~\cite{shamshad2023transformers}. At the same time, the Transformer architecture is prone to ignore local detaile features, which reduces the separability between the background and the foreground of small lesions or objects with large-scale shape changes in medical images. The hybrid architecture of CNN and Transformer realizes both local and global information modeling of medical images by taking the complementary advantages of CNN and Transformer, thus achieving a better medical image segmentation effect. However, these hybrid architectures still suffer from the following two problems. First, these networks ignore the problems of organ deformation and lesion irregularities when modeling local features, resulting in weak local feature expression. Second, these networks ignore the correlation between spacial and channels when modeling the global feature, resulting in inadequate expression of self-attention. To address the above problems, our main contributions are as follows:

\begin{itemize}
  \item A novel dynamic deformable convolution (DDConv) is proposed. Through task adaptive learning, the DDConv can flexibly change the weight coefficient and deformation offset of convolution itself. The DDConv can overcome the problems of fixation of the narrow reception field and non-adaptive convolution kernel parameters existing in the vanilla convolution and its variants, such as Atrous convolution and Involution, etc. The DDConv can improve the ability to perceive tiny lesions and targets with large-scale shape changes in medical images.
  \item A new (shifted)-window adaptive complementary attention module ((S)W-ACAM) is proposed. The (S)W-ACAM realizes the cross-dimensional global modeling of medical images through four parallel branches of weight coefficient adaptive learning. Compared with popular attention mechanisms, such as CBAM and Non-Local, the (S)W-ACAM fully makes up for the deficiency of the conventional attention mechanism in modeling the cross-dimensional relationship between spatial and channels, and thus enhances the separability between the segmented object and the background in medical images.
  \item A new parallel network structure based on dynamically adaptive CNN and cross-dimensional feature fusion Transformer are proposed for medical image segmentation, called TEC-Net. Compared with popular hybrid architectures of CNN and Transformer, like Swin-Unet~\cite{cao2023swin} and Swin UNETR~\cite{tang2022self}. TEC-Net enhances representation learning by tightly combining local and global features at different resolutions through parallel interaction between CNN and Transformer. However, TEC-Net not only abandons pre-training but also requires fewer parameters and fewer computational costs, which are 11.58 M and 4.53 GFLOPs respectively.
\end{itemize}

\section{Related Work}
Medical image segmentation plays a very important role in the field of medical image processing, and is also one of the core techniques of computer-aided diagnosis and treatment systems. Because it’s tedious and complex to label manually medical images, and also it’s difficult to guarantee the efficiency and accuracy of manual labeling, the rapid and accurate segmentation of medical images is of great significance for clinical treatment. In recent years, with the rapid development of deep learning techniques, researchers have continuously developed many deep network models for medical image segmentation. These medical image segmentation networks can be coarsely divided into two categories: CNN and Transformer networks.

\subsection{CNN-based Methods}
Different from traditional medical image segmentation algorithms, the algorithms based on deep learning can learn the high-dimensional feature information of medical images through a multi-layer network structure. Among various deep-learning networks related to medical image segmentation, CNN perform extremely well. CNN can effectively learn from large-scale medical datasets to distinguish features and extract the prior knowledge, making them an important part of smart medical image analysis systems.

In 2015, Ronnebreger et al. were inspired by the FCN~\cite{long2015fully} network and designed the first end-to-end network U-Net~\cite{ronneberger2015u} for medical image segmentation in the ISBI cell tracking challenge. U-Net adopts a symmetric encoder and decoder structure, which can make full use of the local details of medical images and reduce the dependence on training datasets. Therefore, on the case of small datasets, U-Net can still achieve good medical image segmentation results. Based on U-Net, Alom et al. designed R2U-Net~\cite{alom2018recurrent} by combining U-Net, ResNet~\cite{song2020real}, and recurrent neural network (RCNN)~\cite{girshick2014rich}, which has achieved good performance on multiple medical image segmentation datasets such as blood vessels and retinas. To further improve the performance of U-Net, Gu et al. introduced dynamic convolution~\cite{chen2020dynamic} into U-Net and proposed CA-Net~\cite{gu2020net}. Experiments on medical datasets show that CA-Net can not only improve the segmentation accuracy of medical images but also reduce the training time of the network. Inspired by the idea of residual connection and deformable convolution~\cite{dai2017deformable}, Yang et al. added a residual deformable convolution to U-Net, and proposed DCU-Net~\cite{yang2022dcu}. DCU-Net shows a more advanced segmentation effect than U-Net on DRIVE medical dataset. Lei et al. designed SGU-Net~\cite{lei2023sgu} based on U-Net, and proposed an ultralight convolution module and additional adversarial shape-constraint that can significantly improve the segmentation accuracy of abdominal medical images through self-supervised training. Although CNN have made great progress in network structure, the main reason for their success is due to the invariance in dealing with different scales and the inductive bias in local modeling. Although this fixed receptive field improves the computational efficiency of CNN, it also limits the ability to capture the relationship between distant pixels in medical images and lacks the ability to model medical images in a long-range.

\begin{figure*}[htbp]
	\centerline{\includegraphics[width=\textwidth]{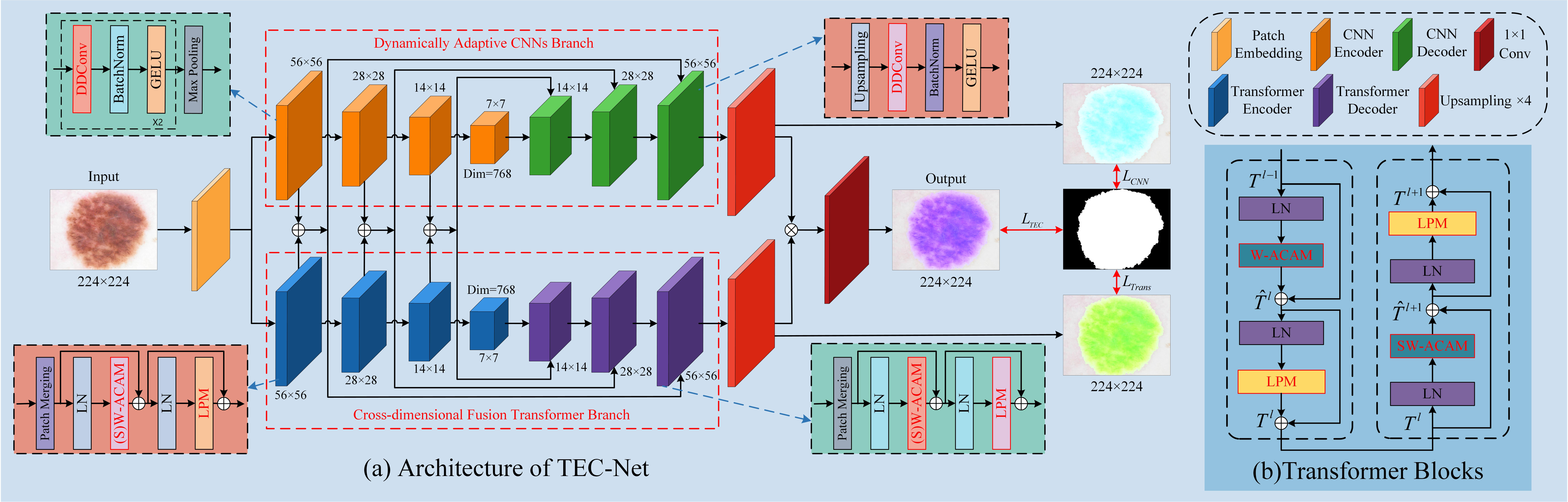}}
	\caption{(a) The architecture of TEC-Net. TEC-Net consists of a dual-branch interaction between dynamically adaptive CNN and cross-dimensional feature fusion Transformer. The DDConv in the CNN branch can adaptively change the weight coefficient and deformation offset of the convolution itself, which improves the segmentation accuracy of irregular objects in medical images. The (S)W-ACAM in the Transformer branch can capture the cross-dimensional long-range dependency in medical images, improving the separability of segmented objects and backgrounds in medical images. The lightweight perceptron module (LPM) greatly reduces the parameters and calculations of the original Transformer network by using the Ghost strategy. (b) Two successive Transformer blocks. W-ACAM and SW-ACAM are cross-dimensional self-attention modules with shifted windows and compact convolutional projection configurations.}
	\label{fig2}
\end{figure*}

\subsection{Transformer-based Methods}
In 2017, Vaswani et al.~\cite{vaswani2017attention} proposed the first Transformer network. Because of its unique structure, Transformer obtains the ability of processing indefinite-length input, establishs long-range dependency relationship, and captures global features of input data. Transformer's success is mainly attributed to the self-attention (SA) mechanism because it can capture long-range dependency.

With the excellent performance of the Transformer in NLP fields, ViT~\cite{dosovitskiy2020image} firstly applies Transformer to the field of image processing, capturing the global context information of input images through multiple cascaded Transformer layers, making Transformer a great success in image classification tasks. Then, Chen et al. proposed TransUNet~\cite{chen2021transunet}, which opened a new era of Transformer in the field of medical image segmentation. As TransUNet directly uses the Transformer network designed for NLP in the field of image segmentation, the size of the input image block is fixed and the calculation is massive. To solve this problem, Valanarasu et al. proposed MedT~\cite{valanarasu2021medical} for medical image segmentation. It adds the gating mechanism to the network, and the gating parameters can be automatically adjusted to obtain the position embedding weight suitable for datasets of different sizes. Since images are more diverse than text and have high resolution, Cao et al. proposed a pure Transformer network Swin-Unet~\cite{cao2023swin} for medical image segmentation by combining the shifted window multi-head self-attention (SW-MSA) in Swin Transformer~\cite{liu2021swin}. Swin-Unet achieved the most advanced segmentation performance at that time on Synapse and ACDC multi-organ segmentation datasets. In order to better use Transformer to process dermoscopic image data, Wang et al. designed BAT~\cite{wang2021boundary} network based on the edge detection idea. The proposed boundary-wise attention gate (BAG) can fully utilize the prior knowledge of image boundaries to capture more details of medical images. BAT achieves amazing segmentation Dice value on the skin lesion datasets and surpasses many of the latest medical image segmentation networks.

Compared with the previous vanilla convolution~\cite{ronneberger2015u}, dynamic convolution~\cite{chen2020dynamic}~\cite{li2021involution} and deformable convolution~\cite{dai2017deformable}, our DDConv can not only adaptively change the weight coefficient and deformation offset of the convolution according to the medical image task, but also better adapt to the shape of organs and small lesions with large-scale changes in medical images, and additionally, it can improve the local feature expression ability of the segmentation network. Compared with the self-attention mechanism in existing Transformer architectures~\cite{cao2023swin}~\cite{wang2021boundary}, our (S)W-ACAM requires fewer parameters and less computation while it is capable of capturing the global cross-dimensional long-range dependency in the medical images, and improving the global feature expression ability of the segmentation network. Our TEC-Net does not require a large amount of labeled data for pre-training, but it can maximize the retention of local details and global semantic information in medical images. It achieves the best segmentation performance on dermoscopic images, liver datasets, and cardiac multi-organ datasets.

\section{Method}

\subsection{Overall Architecture}

\begin{figure}[htbp]
	\centerline{\includegraphics[width=\columnwidth]{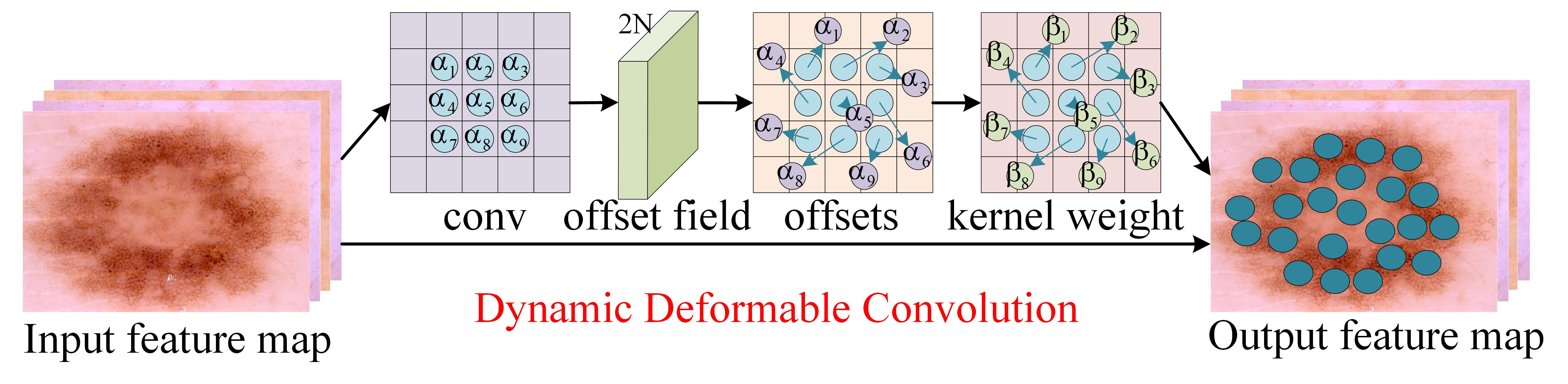}}
	\caption{ The module of the proposed DDConv. Compared with the current popular convolution strategy, DDConv can dynamically adjust the weight coefficient and deformation offset of the convolution itself during the training process, which is conducive to the feature capture and extraction of irregular targets in medical images. $\alpha$ and $\beta$ represent the different weight values of DDConv in different states.}
	\label{fig3}
\end{figure}

The fusion of local and global features are clearly helpful for improving medical image segmentation. CNN capture local features of medical images through convolution operation and hierarchical feature representation. In contrast, the Transformer network realizes the extraction of global features in medical images through the cascaded self-attention mechanism and the matrix operation with context interaction. In order to make full use of local details and global semantic features of medical images, we design a parallel interactive network architecture TEC-Net. The overall architecture of the network is shown in Fig. 1(a). 

TEC-Net fully considers the complementary properties of CNN and Transformer. During the forward propagation process, TEC-Net continuously feeds the local details extracted by the CNN branch to the decoder of the Transformer branch. Similarly, TEC-Net also feeds the global long-range relationships captured by the Transformer branch to the decoder of the CNN branch. Obviously, the proposed TEC-Net provides better local and global feature representation than pure CNN or Transformer networks, and it shows great potential in the field of medical image segmentation.

Specifically, TEC-Net consists of four components: a patch embedding model, a dynamically adaptive CNN branch, a cross-dimensional fusion Transformer branch, and a feature fusion module. Among them, the dynamically adaptive CNN branch and the cross-dimensional fusion Transformer branch follow the design of U-Net and Swin-Unet, respectively. The dynamically adaptive CNN branch consists of seven main stages. By using the weight coefficient and deformation offset adaptive DDConv in each stage, the segmentation network can better understand the local semantic features of medical images, better perceives the subtle changes of human organs or lesions, and improves the ability to extract multi-scale change targets in medical images. Similarly, the cross-dimensional fusion Transformer branch also consists of seven main stages. By using (S)W-ACAM attention in each stage, as shown in Fig. 1(b), the segmentation network can better understand the global dependency of medical images to capture the position information between different organs, and improves the separability of segmented objects and the background in medical images.

Although our TEC-Net can effectively improve the feature representation of medical images, it requires a large number of training data and network parameters due to the dual-branch structure. As the conventional Transformer network contains a lot of multi-layer perceptron (MLP) layers, which not only aggravates the training burden of the network but also causes the number of model parameters rise sharply, resulting in the slow training for the model. Inspired by the idea of the Ghost network~\cite{han2020ghostnet}, we redesign the MLP layer in the original Transformer and proposed a lightweight perceptron module (LPM). The LPM can help our TEC-Net not only achieve better medical image segmentation results than MLP but also greatly reduces the number of parameters and computational costs, even the Transformer can achieve good results without a lot of labeled data training. It is worth mentioning that the dual-branch structure involves mutually symmetric encoders and decoders so that the parallel interaction network structure can maximize the preservation of local features and global features in medical images.

\subsection{Dynamic Deformable Convolution}

Vanilla convolution has spatial invariance and channel specificity, so it has a limited ability to change different visual modalities when dealing with different spatial locations. At the same time, due to the limitations of the receptive field, it is difficult for vanilla convolution to extract features of small targets or targets with blurred edges. Therefore, vanilla convolution inevitably has poor adaptability and weak generalization ability for feature representation of complex medical images. Although the existing deformable convolution~\cite{dai2017deformable} and dynamic convolution~\cite{chen2020dynamic}~\cite{li2021involution} outperforms vanilla convolution to a certain extent, they still face a problem of balancing the performance and size of networks when dealing with medical image segmentation.

In order to overcome the shortcomings of popular convolution operations, this paper proposes a new convolution strategy namely DDConv, as shown in Fig. 2. It can be seen that DDConv can adaptively learn the kernel deformation offset and weight coefficients according to the specific task and data distribution, so as to realize the change of both the shapes and the values of convolution kernels. It can effectively deal with the problems of large data distribution differences and large target deformation in medical image segmentation. Also, DDConv is plug-and-play and can be embedded in any network structure.

The shape change of the convolutional kernel in DDConv is based on the network learning of the deformation offsets. The segmentation network first samples the input feature map $X$ using a square convolutional kernel $S$, and then performs a weighted sum with a weight matrix $M$. The square convolution kernel $S$ determines the range of the receptive field, e.g., a $3 \times 3$ convolution kernel can be expressed as:
\begin{equation}
S=\{( 0,0) ,( 0,1) ,( 0,2) ,...,( 2,1) ,( 2,2)\},
\end{equation}
then the output feature map $Y$ at the coordinate $\varphi_{n}$ can be expressed as:
\begin{equation}
Y\left( \varphi_{n} \right) = {\sum\limits_{\varphi_{m \in S}}{S\left( \varphi_{m} \right)}} \cdot X\left( \varphi_{n} + \varphi_{m} \right),
\end{equation}
when the deformation offset $\bigtriangleup \varphi_{m} = \left\{ m = 1,2,3,\ldots,N \right\}$ is introduced in the weight matrix $M$, $N$ is the total length of $S$. Thus the Equation (2) can be expressed as:
\begin{equation}
Y\left( \varphi_{n} \right) = {\sum\limits_{\varphi_{m \in S}}{S\left( \varphi_{m} \right)}} \cdot X\left( \varphi_{n} + \varphi_{m} + \bigtriangleup \varphi_{m} \right).
\end{equation}

Through network learning, an offset matrix with the same size as the input feature map can be finally obtained, and the matrix dimension is twice that of the input feature map.

To show the convolution kernel of DDConv is dynamic, we first present the output feature map of vanilla convolution:
\begin{equation}
y = \sigma(W \cdot x),
\end{equation}
where $\sigma$ is the activation function, $W$ is the convolutional kernel weight matrix and $y$ is the output feature map. In contrast, the output of the feature map of DDConv is:
\begin{equation}
\hat{y} = \sigma\left( \left( \alpha_{1} \cdot W_{1} + \ldots + \alpha_{n} \cdot W_{n} \right) \cdot x \right),
\end{equation}
where $n$ is the number of weight coefficients, $\alpha_{n}$ is the weight coefficients with learnable parameters and $\hat{y}$ is the output feature map generated by DDConv. DDConv achieves dynamic adjustment of the convolution kernel weights by linearly combining different weight matrices according to the corresponding weight coefficients before performing the convolution operation.

According to the above analysis, we can see that DDConv realizes the dynamic adjustment of the shape and weights of the convolution kernel. Compared with directly increasing the number and size of convolution kernels, the DDConv is simpler and more efficient. The proposed DDConv not only solves the problem of poor adaptive feature extraction ability of fixed-size convolution kernels but also overcomes the defect that different inputs share the same convolution kernel parameters. Consequently, our DDConv can be used to improve the segmentation accuracy of small targets and large targets with blurred edges in medical images.

\begin{table*}[htbp]
\caption{Performance comparison of the proposed method against the SOTA approaches on the ISIC2018 benchmarks. \textcolor{red}{\textbf{Red}} indicates the best result, and \textcolor{blue}{\textbf{blue}} displays the second-best.}
\small
\renewcommand\arraystretch{0.8}
\tabcolsep=0.30cm
\begin{threeparttable}
\begin{tabular}{ccrrrrrrr}
\hline
\multicolumn{2}{c}{\textbf{Method}}                                                & \textbf{DI}$\uparrow$ & \textbf{JA}$\uparrow$ & \textbf{SE}$\uparrow$ & \textbf{AC}$\uparrow$ & \textbf{SP}$\uparrow$ & \textbf{Para. (M) }$\downarrow$ & \textbf{GFLOPs} \\ \hline
\multirow{5}{*}{\textbf{CNN}}                            & U-Net~\cite{ronneberger2015u}           & 86.54         & 79.31          & 88.56          & 93.16          & 96.44           & 34.52                & 65.39           \\
                                                          & R2UNet~\cite{alom2018recurrent}        & 87.92         & 80.28          & 90.92          & 93.38          & 96.33           & 39.09                & 152.82          \\
                                                          & Attention Unet~\cite{oktay2018attention} & 87.16         & 79.55          & 88.52          & 93.17          & 95.62           & 34.88                & 66.57           \\
                                                          & CENet~\cite{gu2019net}          & 87.61         & 81.18          & 90.71          & 94.03          & 96.35           & 29.02                & 11.79           \\
                                                          & CPFNet †~\cite{feng2020cpfnet}       & 90.18         & 82.92          & 91.66          & 94.68          & 96.63           & 30.65                & \textcolor{blue}{\textbf{9.15}}            \\ \hline
\multicolumn{1}{l}{\multirow{6}{*}{\textbf{Transformer}}} & Swin-Unet †~\cite{cao2023swin}     & 89.26         & 80.47          & 90.36          & 94.45          & 96.51           & 41.40                & 11.63           \\
\multicolumn{1}{l}{}                                      & TransUNet †~\cite{chen2021transunet}    & 89.39         & 82.10          & 91.43          & 93.67          & 96.54           & 105.30               & 15.21           \\
\multicolumn{1}{l}{}                                      & BAT †~\cite{wang2021boundary}          & 90.21         & 83.49          & 91.59          & 94.85          & 96.57           & 45.56                & 13.38           \\
\multicolumn{1}{l}{}                                      & CvT †~\cite{wu2021cvt}          & 88.23         & 80.21          & 87.60          & 93.68          & 96.28           & 21.51                & 20.53           \\
\multicolumn{1}{l}{}                                      & PVT~\cite{wang2021pyramid}            & 87.31         & 79.99          & 87.74          & 93.10          & 96.21           & 28.86                & 14.92           \\
\multicolumn{1}{l}{}                                      & CrossForm~\cite{wang2021crossformer}      & 87.44         & 80.06          & 88.25          & 93.39          & 96.40           & 38.66                & 13.57           \\ \hline
\multicolumn{2}{c}{\textbf{TEC-Net-T (our)}}                                       & \textcolor{blue}{\textbf{90.72}}         & \textcolor{blue}{\textbf{84.59}}          & \textcolor{blue}{\textbf{92.54}}          & \textcolor{blue}{\textbf{95.21}}          & \textcolor{blue}{\textbf{96.83}}           & \textcolor{red}{\textbf{11.58}}                & \textcolor{red}{\textbf{4.53}}            \\
\multicolumn{2}{c}{\textbf{TEC-Net-B (our)}}                                       & \textcolor{red}{\textbf{91.23}}         & \textcolor{red}{\textbf{84.76}}          & \textcolor{red}{\textbf{92.68}}          & \textcolor{red}{\textbf{95.56}}          & \textcolor{red}{\textbf{98.21}}           & \textcolor{blue}{\textbf{21.24}}                & 13.29          \\ \hline
\end{tabular}
\begin{tablenotes}
\footnotesize
\item† indicates the model is initialized with pre-trained weights on the ImageNet21K. “Para.” refers to the number of parameters. “GFLOPs” is calculated under the input scale of $224 \times 224$. Since the dermoscopic images are 2D medical images, the comparison methods are all 2D networks.
\end{tablenotes}
\end{threeparttable}
\end{table*}

\subsection{Shifted Window Adaptive Complementary Attention Module}

\begin{figure}[htbp]
	\centerline{\includegraphics[width=\columnwidth]{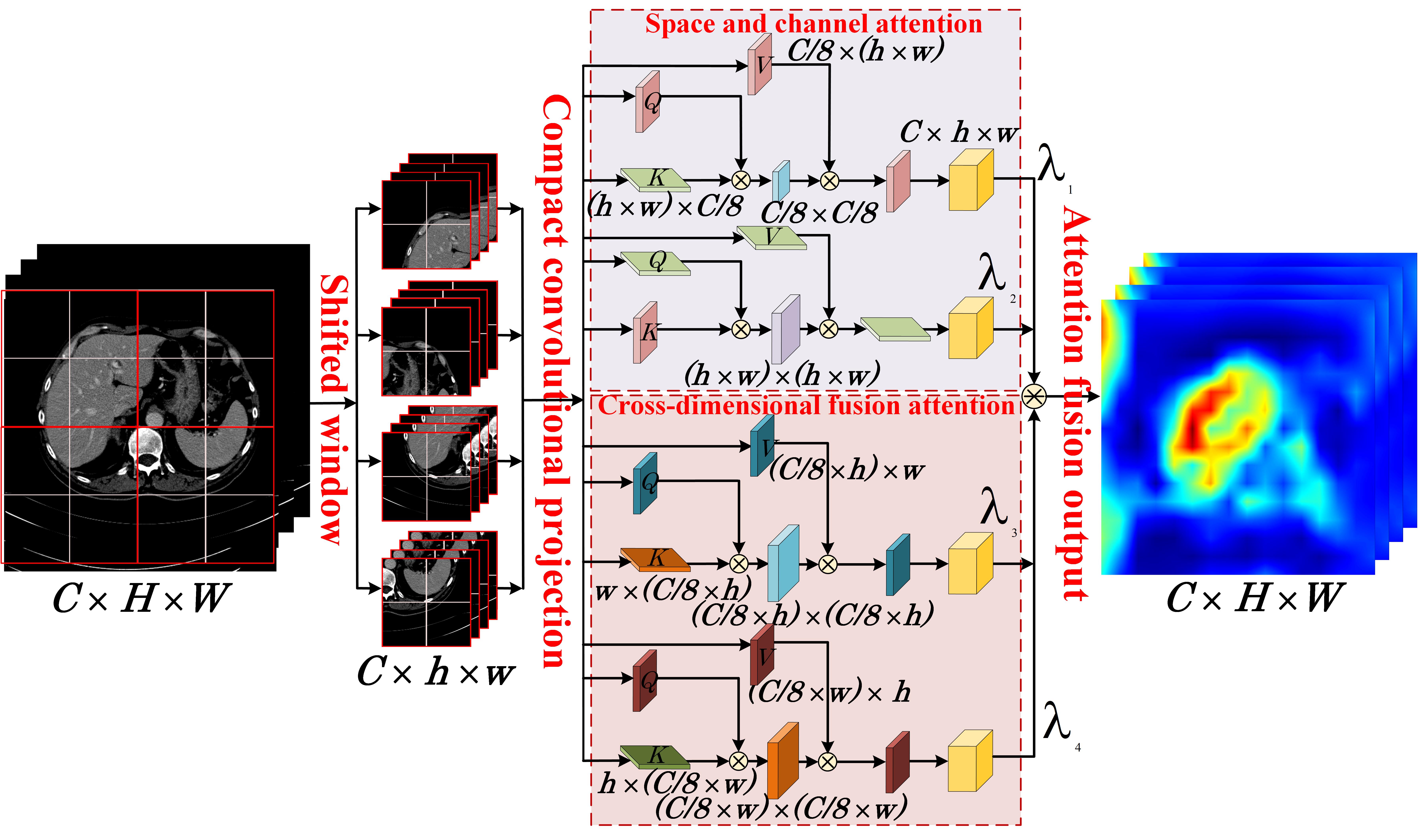}}
	\caption{The module of the proposed (S)W-ACAM. Unlike conventional self-attention, (S)W-ACAM has the advantages of spatial and channel attention, and can also capture long-distance correlation features between spatial and channels. Through the shifted window operation, the spatial resolution of images is significantly reduced, and through the compact convolutional projection operation, the channel dimension of images is also significantly reduced. Thus, the overall computational costs and complexity of our proposed network are reduced. $\lambda_{1}$, $\lambda_{2}$, $\lambda_{3}$ and $\lambda_{4}$ are learnable weight parameters.}
	\label{fig4}
\end{figure}

The self-attention mechanism is the core computing unit in Transformer networks, which realizes the capture of long-range dependency of feature maps by utilizing matrix operations. However, the self-attention mechanism only considers the dependency in the spatial dimension but not the cross-dimensional dependency between spatial and channels~\cite{hong2021qau}. Therefore, when dealing with medical image segmentation with low contrast and high-density noise, the self-attention mechanism is easy to confuse segmentation targets with their background, resulting in poor segmentation results.

To solve the problems mentioned above, we propose a new cross-dimensional self-attention module called (S)W-ACAM. As shown in Fig. 3, (S)W-ACAM consists of four parallel branches, the top two branches are the conventional dual attention module and the bottom two branches are cross-dimensional attention modules. Compared to popular self-attention modules such as spatial self-attention, channel self-attention, and dual self-attention, our proposed (S)W-ACAM can not only fully extract the long-range dependency of both spatial and channels, but also capture the cross-dimensional long-range dependency between spatial and channels. These four branches complement each other, provide richer long-range dependency relationships, enhance the separability between the foreground and background, and thus improve the segmentation results for medical images.

The standard Transformer architecture~\cite{dosovitskiy2020image} uses the global self-attention method to calculate the relationship between one token and all other tokens. This calculation method is complex since the computational costs will increase exponentially with the increase of image size. In order to improve the calculation efficiency, we use the shifted window calculation method similar to that in Swin Transformer~\cite{liu2021swin}, which only calculates the self-attention in the local window. However, in the face of our (S)W-ACAM four branches module, using the shifted window method to calculate self-attention does not reduce the overall computational complexity of the module. Therefore, we also designed the compact convolutional projection. First, we reduce the local size of the medical image through the shifted window operation, then we compress the channel dimension of feature maps through the compact convolutional projection, and finally calculate the self-attention. It is worth mentioning that this method can not only better capture the global information of medical images but also significantly reduce the computational costs of the module. 

Suppose an image contains $h \times w$ windows, each window size is $M \times M$, then the complexity of the (S)W-ACAM, the global MSA in the original Transformer, and the (S)W-MSA in the Swin Transformer are compared as follows:
\begin{equation}
\Omega\left( {MSA} \right) = 4hwC^{2} + 2(hw)^{2}C,
\end{equation}
\begin{equation}
\Omega\left( {(S)W\mbox{-}MSA} \right) = 4hwC^{2} + 2M^{2}hwC,
\end{equation}
\begin{equation}
\Omega\left( {(S)W\mbox{-}ACAM} \right) = \frac{hwC^{2}}{4} + M^{2}hwC,
\end{equation}
if the former term of each formula is a quadratic function of the number of patches $h\cdot w$, the latter term is linear when $M$ is fixed (the default is 7). Then the computational cost of (S)W\mbox{-}ACAM is smaller than MSA and (S)W-MSA.

Among the four parallel branches of (S)W-ACAM, two branches are used to capture channel correlation and spatial correlation, respectively, and the remaining two branches are used to capture the correlation between channel dimension $C$ and space dimension $H$ and vice versa (between channel dimension $C$ and space dimension $W$). After adopting the shifted window partitioning method, as shown in Fig. 1(b), the calculation process of continuous Transformer blocks is as follows:
\begin{equation}
{\hat{T}}^{l} = W\mbox{-}ACAM\left( {LN\left( T^{l - 1} \right)} \right) + T^{l - 1},
\end{equation}
\begin{equation}
T^{l} = LPM\left( {LN\left( {\hat{T}}^{l} \right)} \right) + {\hat{T}}^{l},
\end{equation}
\begin{equation}
{\hat{T}}^{l + 1} = SW\mbox{-}ACAM\left( {LN\left( T^{l} \right)} \right) + T^{l},
\end{equation}
\begin{equation}
T^{l + 1} = LPM\left( {LN\left( {\hat{T}}^{l + 1} \right)} \right) + {\hat{T}}^{l + 1},
\end{equation}
where ${\hat{T}}^{l}$ and $T^{l}$ represent the output features of (S)W-ACAM and LPM, respectively. W-ACAM represents window adaptive complementary attention, SW-ACAM represents shifted window adaptive complementary attention, and LPM represents lightweight perceptron module. For the specific attention calculation process of each branch, we follow the same principle in Swin Transformer as follows:
\begin{equation}
Attention\left( {Q,K,V} \right) = SoftMax\left( {\frac{QK^{T}}{\sqrt{C/8}} + B} \right)V,
\end{equation}
where relative position bias $B\in {{\mathbb{R}}^{{{M}^{2}}\times {{M}^{2}}}}$, $Q,K,V \in \mathbb{R}^{M^{2} \times \frac{C}{8}}$ are query, key, and value matrices respectively. $\frac{C}{8}$ represents the dimension of query/key, and $M^{2}$ represents the number of patches. 

After obtaining ${Out}_{1}$, ${Out}_{2}$, ${Out}_{3}$ and ${Out}_{4}$, the final feature fusion output is:
\begin{equation}
Out = \lambda_{1} \cdot {Out}_{1} + \lambda_{2} \cdot {Out}_{2} + \lambda_{3} \cdot {Out}_{3} + \lambda_{4} \cdot {Out}_{4},
\end{equation}
where $\lambda_{1}$, $\lambda_{2}$, $\lambda_{3}$ and $\lambda_{4}$ are learnable parameters that enable adaptive control of the importance of each attention branch.

Different from other self-attention mechanisms, our proposed (S)W-ACAM can fully capture the correlation between spatial and channels, and reasonably use the context information of medical images to achieve long-range dependency modeling. Since our (S)W-ACAM effectively overcomes the defect that the conventional self-attention only focuses on the spatial self-attention of images and ignores the channel and cross-dimensional self-attention, it achieves better feature representation for medical images with high-density noise, low contrast and complex background.

\begin{table*}[htbp]
\caption{Performance comparison of the proposed method against the SOTA approaches on the LiTS-Liver benchmarks. \textcolor{red}{\textbf{Red}} indicates the best result, and \textcolor{blue}{\textbf{blue}} displays the second-best.}
\small
\renewcommand\arraystretch{0.8}
\tabcolsep=0.13cm
\begin{threeparttable}
\begin{tabular}{lcrrrrrrr}\hline
\multicolumn{2}{c}{\textbf{Method}}                                         & \textbf{DI} $\uparrow$       & \textbf{VOE} $\downarrow$      & \textbf{RVD} $\downarrow$     & \textbf{ASD} $\downarrow$     & \textbf{RMSD} $\downarrow$      & \textbf{Para. (M)} $\downarrow$ & \textbf{GFLOPs} \\ \hline
\multicolumn{1}{c}{\multirow{6}{*}{\textbf{CNN}}} & U-Net~\cite{ronneberger2015u}           & 93.99±1.23 & 11.13±2.47 & 3.22±0.20 & 5.79±0.53 & 123.57±6.28 & 34.52       & 65.39  \\
\multicolumn{1}{c}{}                      & R2UNet~\cite{alom2018recurrent}         & 94.01±1.18 & 11.12±2.37 & 2.36±0.15 & 5.23±0.45 & 120.36±5.03 & 39.09       & 152.82 \\
\multicolumn{1}{c}{}                      & Attention Unet~\cite{oktay2018attention} & 94.08±1.21 & 10.95±2.36 & 3.02±0.18 & 4.95±0.48 & 118.67±5.31 & 34.88       & 66.57  \\
\multicolumn{1}{c}{}                      & CENet~\cite{gu2019net}          & 94.04±1.15 & 11.03±2.31 & 6.19±0.16 & 4.11±0.51 & 115.40±5.82 & 29.02       & \textcolor{blue}{\textbf{11.79}}  \\
\multicolumn{1}{c}{}                      & 3D Unet~\cite{cciccek20163d}         & 94.10±1.06 & 11.13±2.23 & \textcolor{blue}{\textbf{1.42±0.13}} & 2.61±0.45 & 36.43±5.38  & 40.32       & 66.45  \\
\multicolumn{1}{c}{}                      & V-Net~\cite{milletari2016v}           & 94.25±1.03 & 10.65±2.17 & 1.92±0.11 & 2.48±0.38 & 38.28±5.05  & 65.17       & 55.35  \\ \hline
\multirow{5}{*}{\textbf{Transformer}}              & Swin-Unet †~\cite{cao2023swin}     & 95.62±1.32 & 9.73±2.16  & 2.78±0.21 & 2.35±0.35 & 38.85±5.42  & 41.40       & 11.63  \\
                                          & TransUNet †~\cite{chen2021transunet}    & 95.79±1.09 & 9.82±2.10  & 1.98±0.15 & 2.33±0.41 & 37.22±5.23  & 105.30      & 15.21  \\
                                          & CvT †~\cite{wu2021cvt}          & 95.81±1.25 & 9.66±2.31  & 1.77±0.16 & 2.34±0.29 & 36.71±5.09  & 21.51       & 20.53  \\
                                          & PVT~\cite{wang2021pyramid}            & 94.56±1.15 & 9.75±2.19  & 1.69±0.12 & 2.42±0.34 & 37.35±5.16  & 28.86       & 14.92  \\
                                          & CrossForm~\cite{wang2021crossformer}      & 94.63±1.24 & 9.72±2.24  & 1.65±0.15 & 2.39±0.31 & 37.21±5.32  & 38.66       & 13.57  \\ \hline
\multicolumn{2}{c}{\textbf{TEC-Net-T (our)}}                                & \textcolor{blue}{\textbf{96.48±1.05}} & \textcolor{blue}{\textbf{9.53±2.11}}  & 1.45±0.12 & \textcolor{blue}{\textbf{2.29±0.33}} & \textcolor{blue}{\textbf{36.21±4.97}}  & \textcolor{red}{\textbf{11.58}}       & \textcolor{red}{\textbf{4.53}}   \\
\multicolumn{2}{c}{\textbf{TEC-Net-B (our)}}                                & \textcolor{red}{\textbf{96.82±1.22}} & \textcolor{red}{\textbf{9.46±2.33}}  & \textcolor{red}{\textbf{1.38±0.13}} & \textcolor{red}{\textbf{2.21±0.35}} & \textcolor{red}{\textbf{36.08±4.88}}  & \textcolor{blue}{\textbf{21.24}}       & 13.29  \\ \hline
\end{tabular}
\begin{tablenotes}
\footnotesize
\item† indicates the model initialized with pre-trained weights on ImageNet21K. “Para.” refers to the number of parameters. “GFLOPs” is calculated under the input scale of $224 \times 224$. Compared with the comparison experiment on the ISIC2018 dataset, 3D Unet and V-Net are introduced into the comparison experiment on the LiTS-Liver dataset.
\end{tablenotes}
\end{threeparttable}
\end{table*}

\subsection{Loss Function}
In our task, three loss functions are used for model training, namely, the overall loss $L_{TEC}$ of TEC-Net network, the loss $L_{CNN}$ of CNN branch and the loss $L_{Trans}$ of Transformer branch.
\begin{small}
\begin{equation}
L_{TEC} = L_{MSE}\left( {y_{i}^{TEC},y_{i}^{label}} \right) + L_{Dice}\left( {y_{i}^{TEC},y_{i}^{label}} \right),
\end{equation}
\end{small}
\begin{small}
\begin{equation}
L_{CNN} = L_{MSE}\left( {y_{i}^{CNN},y_{i}^{label}} \right) + L_{Dice}\left( {y_{i}^{CNN},y_{i}^{label}} \right),
\end{equation}
\end{small}
\begin{small}
\begin{equation}
L_{Trans} = L_{MSE}\left( {y_{i}^{Trans},y_{i}^{label}} \right) + L_{Dice}\left( {y_{i}^{Trans},y_{i}^{label}} \right),
\end{equation}
\end{small}where $L_{MSE}( \bullet )$ represents mean squared error loss, and $L_{Dice}( \bullet )$ represents Dice loss. $y_{i}^{TEC}$, $y_{i}^{CNN}$, $y_{i}^{Trans}$ and $y_{i}^{label}$ represent the final predicted image using our TEC-Net network, the predicted image from the CNN branch output, the predicted image from the Transformer branch output, and the label, respectively. The total loss function of TEC-Net network can be expressed as:
\begin{small}
\begin{equation}
L_{Total} = \lambda L_{TEC} + \left( {\left( {1 - \lambda} \right)/2} \right)L_{CNN} + \left( {\left( {1 - \lambda} \right)/2} \right)L_{Trans},
\end{equation}
\end{small}where $\lambda = \delta e^{- 5{({1 - k})}^{2}}$, $\lambda$ is a Gaussian ramp-up curve, $k$ represents the number of epochs.

\subsection{Architecture Variants}

We have built a TEC-Net-T as a base network with a model size of 11.58 M and a computing capacity of 4.53 GFLOPs. In addition, we built the TEC-Net-B network to make a fair comparison with the latest networks such as CvT~\cite{wu2021cvt} and PVT~\cite{wang2021pyramid}. The window size is set to 7, and the input image size is $224 \times 224$. Other network parameters are set as follows:

\begin{itemize}
  \item TEC-Net-T: $layer~number = \left\{ 2,~2,~6,~2,~6,~2,~2 \right\}$, $H = \left\{ 3,~6,~12,~24,~12,~6,~3 \right\}$, $D = 96$
  \item TEC-Net-B: $layer~number = \left\{ 2,~2,~18,~2,~18,~2,~2 \right\}$, $H = \left\{ 4,~8,~16,~32,~16,~8,~4 \right\}$, $D = 96$ 
\end{itemize}
where $D$ represents the number of image channels when entering the first layer of the dynamically adaptive CNN branch and the cross-dimensional fusion Transformer branch, $layer ~number$ represents the number of Transformer blocks used in each stage, and $H$ represents the number of multiple heads in the Transformer branch.

\section{Experiment and Results}
\subsection{Datasets}
We conducted experiments on the skin lesion segmentation dataset (ISIC2018) from the International Symposium on Biomedical Imaging (ISBI)~\cite{codella2019skin}, the Liver Tumor Segmentation Challenge dataset (LiTS) from the Medical Image Computing and Computer Assisted Intervention Society (MICCAI)~\cite{bilic2023liver} and the Automated Cardiac Diagnosis Challenge dataset (ACDC) from the University Hospital of Dijon (France)~\cite{bernard2018deep}. These three datasets have different data types and data distributions. Among them, the ISIC2018 dataset is electron microscope images, which contain 2,594 dermoscopic images for training, but the ground truth images of the testing set have not been released, so we performed a five-fold cross-validation on the training set for a fair comparison. The LiTS dataset is a CT image of the human abdomen, containing 131 3D CT liver scans, where 100 scans of which are used for training, and the remaining 31 scans are used for testing. The ACDC dataset is an MRI image of the human heart, which contains cardiac short-axis Cine MRI data from 100 patients. The ACDC dataset includes healthy patients, patients with previous myocardial infarction, dilated cardiomyopathy, hypertrophic cardiomyopathy, and abnormal right ventricle, with 20 scans for each group. These data are obtained over a 6 years period using two MRI scanners of two magnetic strengths (1.5T and 3.0T). In addition, all images are empirically resized to 224×224 for efficiency.

\subsection{Implementation Details and Evaluation Indicators}
All the networks are implemented on NVIDIA GeForce RTX 3090 24GB and PyTorch 1.7. We utilized Adam with an initial learning rate of 0.001 to optimize the networks. The learning rate decreases in half when the loss on the validation set has not dropped by 10 epochs. We used mean squared error loss (MSE) and Dice loss as loss functions in our experiment.

In the experiment on the ISIC2018 dataset, we conducted an overall evaluation for SOTA networks and the proposed TEC-Net using five indicators: Dice (DI), Jaccard (JA), Sensitivity (SE), Accuracy (AC), and Specificity (SP)~\cite{chang2009performance}. In the experiment on the LiTS-Liver dataset, we conducted an overall evaluation for SOTA networks and the proposed TEC-Net using five indicators: DI, VOE, RVD, ASD, and RMSD. In the experiment on the ACDC dataset, we conducted an overall evaluation for SOTA networks and the proposed TEC-Net using DI and 95HD~\cite{taha2015metrics}.

\begin{table*}[htbp]
\caption{Performance comparison of the proposed method against the SOTA approaches on the ACDC benchmarks. \textcolor{red}{\textbf{Red}} indicates the best result, and \textcolor{blue}{\textbf{blue}} displays the second-best.}
\small
\renewcommand\arraystretch{0.9}
\tabcolsep=0.06cm
\begin{threeparttable}
\begin{tabular}{ccrrrrrrrr}
\hline
\multicolumn{2}{c}{\multirow{2}{*}{\textbf{Model}}}            & \multicolumn{2}{c}{\textbf{RV}}                            & \multicolumn{2}{c}{\textbf{MYO}}                           & \multicolumn{2}{c}{\textbf{LV}}                            & \multicolumn{2}{c}{\textbf{AVG}}                           \\
\multicolumn{2}{c}{}                                  & \multicolumn{1}{c}{DI} & \multicolumn{1}{c}{95HD} & \multicolumn{1}{c}{DI} & \multicolumn{1}{c}{95HD} & \multicolumn{1}{c}{DI} & \multicolumn{1}{c}{95HD} & \multicolumn{1}{c}{DI} & \multicolumn{1}{c}{95HD} \\ \hline
\multirow{6}{*}{\textbf{CNN}}        & U-Net~\cite{ronneberger2015u}           & 90.11±1.20             & 5.90±1.36                & 88.87±0.33             & 2.49±0.51                & 94.16±1.06             & 2.95±0.77                & 91.06±0.68             & 3.79±0.75                \\
                             & R2UNet~\cite{alom2018recurrent}         & 90.03±1.19             & 6.92±2.30                & 89.15±0.42             & 2.46±0.32                & 94.33±1.04             & \textcolor{blue}{\textbf{2.85±0.46}}                & 91.21±0.72             & 4.06±0.77                \\
                             & Attention Unet~\cite{oktay2018attention} & 90.41±1.06             & 5.48±0.93                & 88.94±0.36             & 2.81±0.78                & 94.14±1.37             & 3.26±1.32                & 91.18±0.82             & 3.85±0.93                \\
                             & CENet~\cite{gu2019net}          & 89.77±0.28             & 5.32±0.29                & 88.95±0.23             & 2.53±0.28                & 94.16±1.33             & 2.93±1.12                & 91.29±0.22             & 3.69±0.81                \\
                             & 3D Unet~\cite{cciccek20163d}         & 90.25±0.82             & \textcolor{blue}{\textbf{4.95±1.02}}                & 89.07±0.41             & 2.51±0.33                & 94.23±1.40             & 3.15±1.09                & 91.33±0.47             & 3.71±0.69                \\
                             & V-Net~\cite{milletari2016v}           & 90.01±0.76             & 5.05±0.88                & 89.00±0.29             & 2.47±0.46                & 94.13±1.31             & 3.24±1.43                & 91.30±0.72             & 3.64±0.83                \\ \hline
\multirow{5}{*}{\textbf{Transformer}} & Swin-Unet †~\cite{cao2023swin}    & 90.35±1.13             & 5.45±1.08                & 88.73±0.49             & 3.16±1.32                & 94.03±1.43             & 3.77±2.23                & 91.04±0.93             & 4.13±1.51                \\
                             & TransUNet †~\cite{chen2021transunet}    & 89.79±1.28             & 6.35±1.31                & 88.77±0.56             & 3.09±1.30                & 94.05±1.27             & 3.38±1.55                & 90.84±0.98             & 4.27±1.26                \\
                             & CvT †~\cite{wu2021cvt}          & 90.05±1.36             & 5.82±1.43                & 88.74±0.39             & 3.11±0.98                & 94.06±1.36             & 3.45±1.65                & 90.97±0.86             & 4.17±1.34                \\
                             & PVT~\cite{wang2021pyramid}            & 89.52±1.31             & 6.51±1.35                & 88.08±0.52             & 3.12±1.35                & 93.75±1.25             & 3.34±1.53                & 90.46±0.78             & 4.23±1.14                \\
                             & CrossForm~\cite{wang2021crossformer}      & 89.66±1.48             & 6.23±1.51                & 88.17±0.49             & 3.08±1.27                & 93.97±1.06             & 3.28±1.42                & 90.57±0.74             & 4.19±1.08                \\ \hline
\multicolumn{2}{c}{\textbf{TEC-Net-T (our)}}                   & \textcolor{blue}{\textbf{90.46±1.15}}             & 5.27±0.55                & \textcolor{blue}{\textbf{89.33±0.48}}             & \textcolor{blue}{\textbf{2.43±0.46}}                & \textcolor{blue}{\textbf{94.38±1.23}}             & 2.87±0.81                & \textcolor{blue}{\textbf{91.42±0.82}}             & \textcolor{blue}{\textbf{3.55±0.54}}                \\
\multicolumn{2}{c}{\textbf{TEC-Net-B (our)}}                   & \textcolor{red}{\textbf{90.95±0.65}}             & \textcolor{red}{\textbf{4.86±0.37}}                & \textcolor{red}{\textbf{89.62±0.36}}             & \textcolor{red}{\textbf{2.32±0.82}}                & \textcolor{red}{\textbf{95.30±1.08}}             & \textcolor{red}{\textbf{2.63±0.83}}                & \textcolor{red}{\textbf{91.95±0.66}}             & \textcolor{red}{\textbf{3.45±0.51}}                \\ \hline       
\end{tabular}
\begin{tablenotes}
\footnotesize
\item† indicates the model initialized with pre-trained weights on ImageNet21K.
\end{tablenotes}
\end{threeparttable}
\end{table*}

The values of DI, JA, AC, SE, and SP range from 0 to 100, and better segmentation results mean that the values of evaluation indicators such as DI, JA, AC, SE, and SP will be higher, while the values of evaluation indicators such as VOE, RVD, ASD, RMSD, and 95HD will be lower~\cite{li2020transformation}.The 95HD is defined as the 95th quantile of Hausdorff distances (HD) instead of the maximum~\cite{lei2022semi}.
\subsection{Evaluation and Results}
In this paper, we selected popular SOTA networks for medical image segmentation networks U-Net~\cite{ronneberger2015u}, R2UNet~\cite{alom2018recurrent}, Attention Unet~\cite{oktay2018attention}, CENet~\cite{gu2019net}, 3D Unet~\cite{cciccek20163d}, V-Net~\cite{milletari2016v}, Swin-Unet~\cite{cao2023swin}, TransUNet~\cite{chen2021transunet}, CvT~\cite{wu2021cvt}, PVT~\cite{wang2021pyramid}, CrossForm~\cite{wang2021crossformer} and the proposed TEC-Net to conduct a comprehensive comparison of the three different modalities datasets, the ISIC2018, the LiTS-Liver and the ACDC.

TABLE I shows the quantitative analysis results of the proposed TEC-Net and the competitive CNN and Transformer networks on the ISIC2018 dataset. From the experimental results, we can conclude that our TEC-Net needs the minimum number of parameters and the lowest computational costs, and can obtain the best segmentation effect on the dermoscopic images without adding pre-training. Moreover, our TEC-Net-T network requires only 11.58 M of parameters and 4.53 GFLOPs of computational costs, but still achieves the second-best segmentation effect. Our TEC-Net-B network, BAT, CvT, and CrossForm have similar parameters or computational costs, but on the ISIC2018 dataset, the division Dice value of our TEC-Net-B is 1.02\%, 3.00\%, and 3.79\% higher than that of the BAT, CvT, and CrossForm network respectively. In terms of other evaluation indicators, our TEC-Net-B is also significantly better than other competitive networks.

TABLE II shows the quantitative analysis results of the proposed TEC-Net and the competitive networks on the LiTS-Liver dataset. It can be seen from the experimental results that our TEC-Net shows great advantages in medical image segmentation, which further verifies the integrity of TEC-Net in extracting local and global features for medical images. It is worth noting that the TEC-Net-B and TEC-Net-T networks achieve good results in medical image segmentation in the first and second place, with the least number of model parameters and computational costs. The division Dice value of our TEC-Net-B network without pre-training is 1.20\%, 1.03\%, and 1.01\% higher than that of the Swin-Unet, TransUNet, and CvT network with pre-training. In terms of other evaluation indicators, our TEC-Net-B is also significantly better than other competitive networks.

\begin{figure}[htbp]
\centerline{\includegraphics[width=0.8\columnwidth]{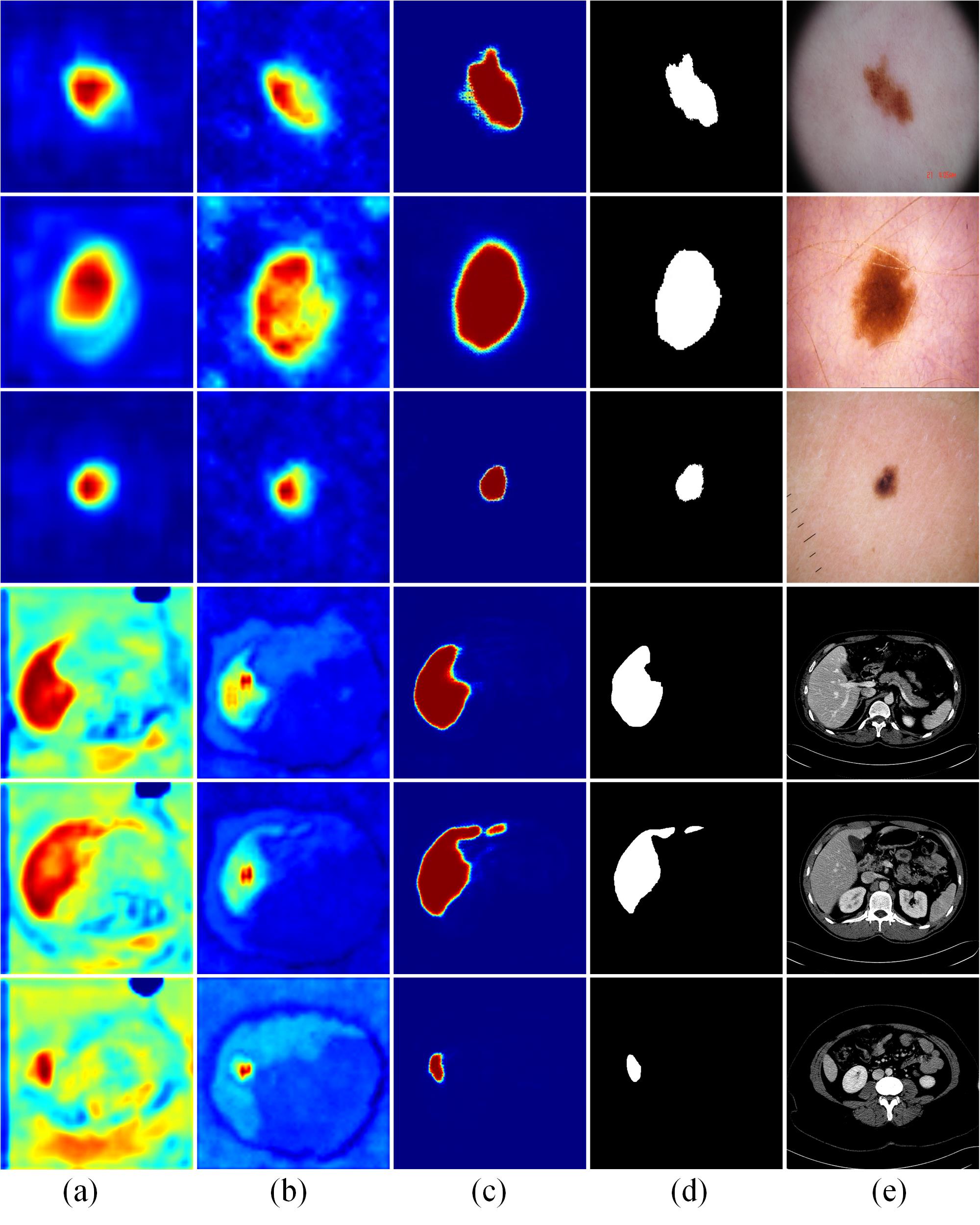}}
\caption{Visualization effect display of TEC-Net network on ISIC2018 dataset and LiTS-Liver dataset. (a) The predicted image output by the CNN branch, (b) the predicted image output by the Transformer branch, (c) the predicted image output by the TEC-Net network, (d) the corresponding label, (e) the corresponding original image.}
\label{fig5}
\end{figure}

TABLE III shows the quantitative analysis results of the proposed TEC-Net and the competitive networks on the ACDC dataset. From the experimental results, it can be seen that the proposed TEC-Net still exhibits significant advantages on MRI type multi-organ segmentation datasets. Both TEC-Net-T and TEC-Net-B provide state-of-the-art segmentation effects for the left ventricle (LV), right ventricle (RV), and left ventricular myocardium (MYO). Among them, LV provides the best segmentation effect, while MYO provides a poor segmentation effect. Compared with the latest CvT, PVT, and CrossForm, the average segmentation performance of TEC-Net-B improves by 0.98\%, 1.45\%, and 1.38\%, respectively, while the average 95HD decreases by 0.72\%, 0.78\%, and 0.74\%, respectively. It also proves that TEC-Net has strong generalization performance on different datasets and can be flexibly applied to medical image segmentation tasks in different modalities and collection environments.

From the visualization in Fig. 4, it can be clearly seen that the TEC-Net network can effectively extract local detail features and global semantic features in medical images to the maximum extent possible. Therefore, it provides accurate segmentation results for segmentation targets with irregular edges and large deformation scales. Fig. 4(a) shows the final output result from the CNN branch, and Fig. 4(b) shows the final output result from the Transformer branch. We can find that the CNN branch captures more accurate local detail information of segmented targets. However, due to the fact that CNN branches mainly capture local features of images, they are more susceptible to noise interference. However, the Transformer branch captures more accurate position information of segmented targets. Due to the fact that the Transformer branch mainly captures the global features of the image, it is easy to overlook the detailed information of the segmented target. So, after integrating the CNN branch with the Transformer branch, the TEC-Net network fully inherits the structural and generalization advantages of CNN and Transformer, providing better local and global feature representations for medical images, demonstrating great potential in the field of medical image segmentation.
\subsection{Ablation Study}
In order to fully prove the effectiveness of different modules in our TEC-Net, we conducted a series of ablation experiments on the ISIC2018 dataset. As shown in TABLE IV, we can see that the DDConv and (S)W-ACAM proposed in this paper show good performance, and the combination of these two modules, TEC-Net shows the best medical image segmentation effect.

\begin{table}[htbp]
\caption{Ablation experiments of DDConv, (S)W-ACAM and LPM in TEC-Net on the ISIC2018 dataset.}
\footnotesize
\renewcommand\arraystretch{0.95}
\tabcolsep=0.02cm
\begin{tabular}{ccccrr}
\hline
\multicolumn{1}{c}{\textbf{Backbone}} & \multicolumn{1}{c}{\textbf{DDConv}} & \multicolumn{1}{c}{\textbf{(S)W-ACAM}} & \multicolumn{1}{c}{\textbf{LPM}} & \multicolumn{1}{c}{\textbf{Para. (M)}} & \multicolumn{1}{c}{\textbf{DI (\%) $\uparrow$}} \\ \hline
U-Net+Swin-Unet              &                            &                             &                         & 46.92                         & 87.45                         \\
U-Net+Swin-Unet              & $\surd$                          &                             &                         & 48.25                         & 89.15                         \\
U-Net+Swin-Unet              &                            & $\surd$                           &                         & 30.26                         & 89.62                         \\
U-Net+Swin-Unet              &                            &                             & $\surd$                       & 15.45                         & 88.43                         \\
U-Net+Swin-Unet              & $\surd$                          & $\surd$                           &                         & 32.16                         & 90.88                         \\
U-Net+Swin-Unet              & $\surd$                    &                                   & $\surd$                 & 16.93                         & 89.12                        \\
U-Net+Swin-Unet              &                            & $\surd$                           & $\surd$                       & 9.67                          & 89.46                         \\
U-Net+Swin-Unet              & $\surd$                    & C-C                               & $\surd$                       & 10.69                         & 88.82             \\
U-Net+Swin-Unet              & $\surd$                    & C-H                               & $\surd$                       & 10.24                         & 89.05             \\
U-Net+Swin-Unet              & $\surd$                    & C-W                               & $\surd$                       & 10.24                         & 89.06             \\
U-Net+Swin-Unet              & $\surd$                    & H-W                               & $\surd$                       & 10.05                         & 88.95             \\
\textbf{TEC-Net-T (our)}              & $\surd$                    & $\surd$                           & $\surd$                       & 11.58                         & 90.72               \\ \hline  
\end{tabular}
\end{table}

\subsection{Network Visualization}

In order to better understand the TEC-Net network, we visualized the feature maps of each stage of the network, as shown in Fig. 5. We used the linear interpolation method to restore the deep feature map with the low spatial resolution to the same size as the input and output images. In the visualization, red represents areas that the network is more concerned about, while blue represents areas that the network is less concerned about.

Through visualization images, we can clearly see the entire process of feature extraction, feature fusion, and other stages that the image undergoes after being input into the network. In the two branches of the encoder, the model can analyze and identify the semantic information of images. In the two branches of the decoder, after skip connections and feature fusion operations, the model began to pay more attention to the semantic features of target regions. During this process, the information exchange and fusion between CNN and Transformer branches play an important role in the accurate segmentation of target regions.

As a whole structure, as the number of network layers continues to deepen, the TEC-Net network gradually refines the localization and contour segmentation of semantic objects in medical images, proving the effectiveness of our proposed TEC-Net network for global information modeling and accurate segmentation of medical images.

\begin{figure}[htbp]
	\centerline{\includegraphics[width=\columnwidth]{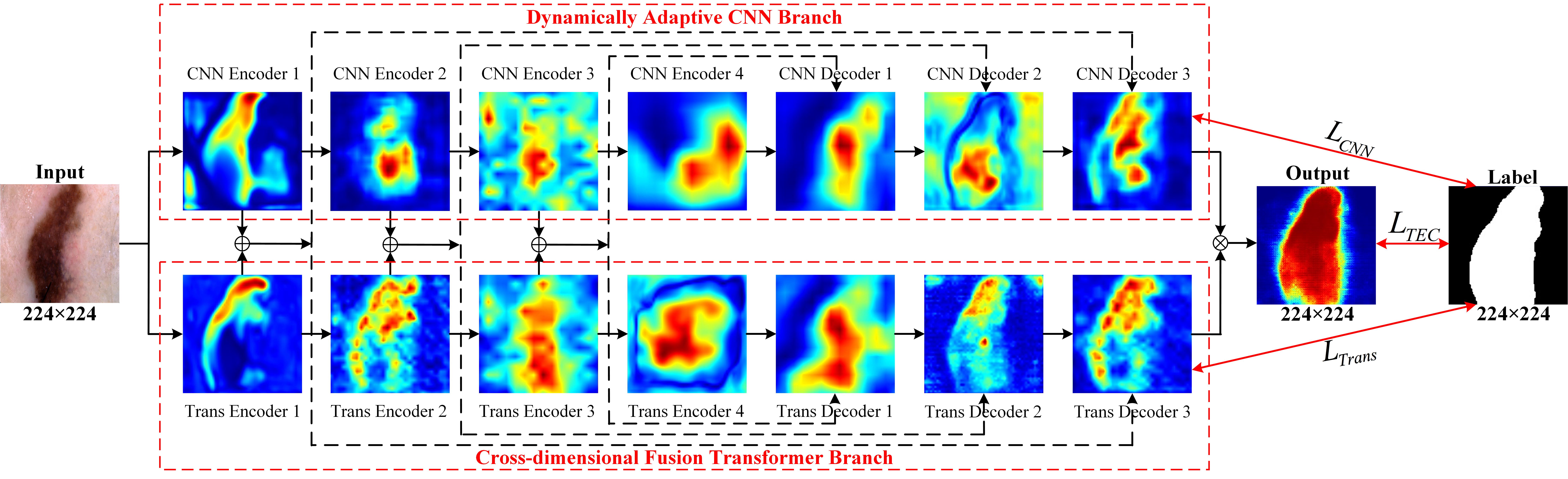}}
	\caption{Visualization of results at each layer of TEC-Net (The original image is from the ISIC2018 dataset).}
	\label{fig6}
\end{figure}

\section{Conclusion}
In this study, we have proposed a new architecture TEC-Net that combined dynamically adaptive CNN and cross-dimensional fusion Transformer in parallel for medical image segmentation. The proposed TEC-Net integrates the advantages of both CNN and Transformer, and retains the local details and global semantic features of medical images using local relationship modeling and long-range dependency modeling. The proposed DDConv overcomes the problems of fixed receptive field and parameter sharing in vanilla convolution, enhances the ability to express local features, and realizes adaptive extraction of spatial features. The proposed (S)W-ACAM self-attention mechanism can fully capture the cross-dimensional correlation between spatial and channels, and adaptively learn the important information between spatial and channels through network training. In addition, by using the LPM to replace the MLP in the traditional Transformer, our TEC-Net significantly reduces the number of parameters, gets rid of the dependence of the network on pre-training, avoids the challenge of lacking labeled medical images and easy suffering from over-fitting. Compared with popular CNN and Transformer medical image segmentation networks, our TEC-Net shows significant advantages in terms of operational efficiency and segmentation effect.

\bibliographystyle{named}
\bibliography{ijcai23}

\begin{thebibliography}{}

\bibitem[\protect\citeauthoryear{Alom \bgroup \em et al.\egroup
  }{2018}]{alom2018recurrent}
Md~Zahangir Alom, Mahmudul Hasan, Chris Yakopcic, Tarek~M Taha, and Vijayan~K
  Asari.
\newblock Recurrent residual convolutional neural network based on u-net
  (r2u-net) for medical image segmentation.
\newblock {\em arXiv preprint arXiv:1802.06955}, 2018.

\bibitem[\protect\citeauthoryear{Bernard \bgroup \em et al.\egroup
  }{2018}]{bernard2018deep}
Olivier Bernard, Alain Lalande, Clement Zotti, Frederick Cervenansky, Xin Yang,
  Pheng-Ann Heng, Irem Cetin, Karim Lekadir, Oscar Camara, Miguel
  Angel~Gonzalez Ballester, et~al.
\newblock Deep learning techniques for automatic mri cardiac multi-structures
  segmentation and diagnosis: is the problem solved?
\newblock {\em IEEE transactions on medical imaging}, 37(11):2514--2525, 2018.

\bibitem[\protect\citeauthoryear{Bilic \bgroup \em et al.\egroup
  }{2023}]{bilic2023liver}
Patrick Bilic, Patrick Christ, Hongwei~Bran Li, Eugene Vorontsov, Avi
  Ben-Cohen, Georgios Kaissis, Adi Szeskin, Colin Jacobs, Gabriel
  Efrain~Humpire Mamani, Gabriel Chartrand, et~al.
\newblock The liver tumor segmentation benchmark (lits).
\newblock {\em Medical Image Analysis}, 84:102680, 2023.

\bibitem[\protect\citeauthoryear{Cao \bgroup \em et al.\egroup
  }{2023}]{cao2023swin}
Hu~Cao, Yueyue Wang, Joy Chen, Dongsheng Jiang, Xiaopeng Zhang, Qi~Tian, and
  Manning Wang.
\newblock Swin-unet: Unet-like pure transformer for medical image segmentation.
\newblock In {\em Computer Vision--ECCV 2022 Workshops: Tel Aviv, Israel,
  October 23--27, 2022, Proceedings, Part III}, pages 205--218. Springer, 2023.

\bibitem[\protect\citeauthoryear{Chang \bgroup \em et al.\egroup
  }{2009}]{chang2009performance}
Herng-Hua Chang, Audrey~H Zhuang, Daniel~J Valentino, and Woei-Chyn Chu.
\newblock Performance measure characterization for evaluating neuroimage
  segmentation algorithms.
\newblock {\em Neuroimage}, 47(1):122--135, 2009.

\bibitem[\protect\citeauthoryear{Chen \bgroup \em et al.\egroup
  }{2020}]{chen2020dynamic}
Yinpeng Chen, Xiyang Dai, Mengchen Liu, Dongdong Chen, Lu~Yuan, and Zicheng
  Liu.
\newblock Dynamic convolution: Attention over convolution kernels.
\newblock In {\em Proceedings of the IEEE/CVF conference on computer vision and
  pattern recognition}, pages 11030--11039, 2020.

\bibitem[\protect\citeauthoryear{Chen \bgroup \em et al.\egroup
  }{2021}]{chen2021transunet}
Jieneng Chen, Yongyi Lu, Qihang Yu, Xiangde Luo, Ehsan Adeli, Yan Wang, Le~Lu,
  Alan~L Yuille, and Yuyin Zhou.
\newblock Transunet: Transformers make strong encoders for medical image
  segmentation.
\newblock {\em arXiv preprint arXiv:2102.04306}, 2021.

\bibitem[\protect\citeauthoryear{{\c{C}}i{\c{c}}ek \bgroup \em et al.\egroup
  }{2016}]{cciccek20163d}
{\"O}zg{\"u}n {\c{C}}i{\c{c}}ek, Ahmed Abdulkadir, Soeren~S Lienkamp, Thomas
  Brox, and Olaf Ronneberger.
\newblock 3d u-net: learning dense volumetric segmentation from sparse
  annotation.
\newblock In {\em Medical Image Computing and Computer-Assisted
  Intervention--MICCAI 2016: 19th International Conference, Athens, Greece,
  October 17-21, 2016, Proceedings, Part II 19}, pages 424--432. Springer,
  2016.

\bibitem[\protect\citeauthoryear{Codella \bgroup \em et al.\egroup
  }{2019}]{codella2019skin}
Noel Codella, Veronica Rotemberg, Philipp Tschandl, M~Emre Celebi, Stephen
  Dusza, David Gutman, Brian Helba, Aadi Kalloo, Konstantinos Liopyris, Michael
  Marchetti, et~al.
\newblock Skin lesion analysis toward melanoma detection 2018: A challenge
  hosted by the international skin imaging collaboration (isic).
\newblock {\em arXiv preprint arXiv:1902.03368}, 2019.

\bibitem[\protect\citeauthoryear{Dai \bgroup \em et al.\egroup
  }{2017}]{dai2017deformable}
Jifeng Dai, Haozhi Qi, Yuwen Xiong, Yi~Li, Guodong Zhang, Han Hu, and Yichen
  Wei.
\newblock Deformable convolutional networks.
\newblock In {\em Proceedings of the IEEE international conference on computer
  vision}, pages 764--773, 2017.

\bibitem[\protect\citeauthoryear{Dosovitskiy \bgroup \em et al.\egroup
  }{2020}]{dosovitskiy2020image}
Alexey Dosovitskiy, Lucas Beyer, Alexander Kolesnikov, Dirk Weissenborn,
  Xiaohua Zhai, Thomas Unterthiner, Mostafa Dehghani, Matthias Minderer, Georg
  Heigold, Sylvain Gelly, et~al.
\newblock An image is worth 16x16 words: Transformers for image recognition at
  scale.
\newblock {\em arXiv preprint arXiv:2010.11929}, 2020.

\bibitem[\protect\citeauthoryear{Feng \bgroup \em et al.\egroup
  }{2020}]{feng2020cpfnet}
Shuanglang Feng, Heming Zhao, Fei Shi, Xuena Cheng, Meng Wang, Yuhui Ma, Dehui
  Xiang, Weifang Zhu, and Xinjian Chen.
\newblock Cpfnet: Context pyramid fusion network for medical image
  segmentation.
\newblock {\em IEEE transactions on medical imaging}, 39(10):3008--3018, 2020.

\bibitem[\protect\citeauthoryear{Girshick \bgroup \em et al.\egroup
  }{2014}]{girshick2014rich}
Ross Girshick, Jeff Donahue, Trevor Darrell, and Jitendra Malik.
\newblock Rich feature hierarchies for accurate object detection and semantic
  segmentation.
\newblock In {\em Proceedings of the IEEE conference on computer vision and
  pattern recognition}, pages 580--587, 2014.

\bibitem[\protect\citeauthoryear{Gu \bgroup \em et al.\egroup
  }{2019}]{gu2019net}
Zaiwang Gu, Jun Cheng, Huazhu Fu, Kang Zhou, Huaying Hao, Yitian Zhao, Tianyang
  Zhang, Shenghua Gao, and Jiang Liu.
\newblock Ce-net: Context encoder network for 2d medical image segmentation.
\newblock {\em IEEE transactions on medical imaging}, 38(10):2281--2292, 2019.

\bibitem[\protect\citeauthoryear{Gu \bgroup \em et al.\egroup
  }{2020}]{gu2020net}
Ran Gu, Guotai Wang, Tao Song, Rui Huang, Michael Aertsen, Jan Deprest,
  S{\'e}bastien Ourselin, Tom Vercauteren, and Shaoting Zhang.
\newblock Ca-net: Comprehensive attention convolutional neural networks for
  explainable medical image segmentation.
\newblock {\em IEEE transactions on medical imaging}, 40(2):699--711, 2020.

\bibitem[\protect\citeauthoryear{Han \bgroup \em et al.\egroup
  }{2020}]{han2020ghostnet}
Kai Han, Yunhe Wang, Qi~Tian, Jianyuan Guo, Chunjing Xu, and Chang Xu.
\newblock Ghostnet: More features from cheap operations.
\newblock In {\em Proceedings of the IEEE/CVF conference on computer vision and
  pattern recognition}, pages 1580--1589, 2020.

\bibitem[\protect\citeauthoryear{Hong \bgroup \em et al.\egroup
  }{2021}]{hong2021qau}
Luminzi Hong, Risheng Wang, Tao Lei, Xiaogang Du, and Yong Wan.
\newblock Qau-net: Quartet attention u-net for liver and liver-tumor
  segmentation.
\newblock In {\em 2021 IEEE International Conference on Multimedia and Expo
  (ICME)}, pages 1--6. IEEE, 2021.

\bibitem[\protect\citeauthoryear{Lei \bgroup \em et al.\egroup
  }{2022}]{lei2022semi}
Tao Lei, Dong Zhang, Xiaogang Du, Xuan Wang, Yong Wan, and Asoke~K Nandi.
\newblock Semi-supervised medical image segmentation using adversarial
  consistency learning and dynamic convolution network.
\newblock {\em IEEE Transactions on Medical Imaging}, 2022.

\bibitem[\protect\citeauthoryear{Lei \bgroup \em et al.\egroup
  }{2023}]{lei2023sgu}
Tao Lei, Rui Sun, Xiaogang Du, Huazhu Fu, Changqing Zhang, and Asoke~K Nandi.
\newblock Sgu-net: Shape-guided ultralight network for abdominal image
  segmentation.
\newblock {\em IEEE Journal of Biomedical and Health Informatics}, 2023.

\bibitem[\protect\citeauthoryear{Li \bgroup \em et al.\egroup
  }{2020}]{li2020transformation}
Xiaomeng Li, Lequan Yu, Hao Chen, Chi-Wing Fu, Lei Xing, and Pheng-Ann Heng.
\newblock Transformation-consistent self-ensembling model for semisupervised
  medical image segmentation.
\newblock {\em IEEE Transactions on Neural Networks and Learning Systems},
  32(2):523--534, 2020.

\bibitem[\protect\citeauthoryear{Li \bgroup \em et al.\egroup
  }{2021}]{li2021involution}
Duo Li, Jie Hu, Changhu Wang, Xiangtai Li, Qi~She, Lei Zhu, Tong Zhang, and
  Qifeng Chen.
\newblock Involution: Inverting the inherence of convolution for visual
  recognition.
\newblock In {\em Proceedings of the IEEE/CVF Conference on Computer Vision and
  Pattern Recognition}, pages 12321--12330, 2021.

\bibitem[\protect\citeauthoryear{Liu \bgroup \em et al.\egroup
  }{2021}]{liu2021swin}
Ze~Liu, Yutong Lin, Yue Cao, Han Hu, Yixuan Wei, Zheng Zhang, Stephen Lin, and
  Baining Guo.
\newblock Swin transformer: Hierarchical vision transformer using shifted
  windows.
\newblock In {\em Proceedings of the IEEE/CVF international conference on
  computer vision}, pages 10012--10022, 2021.

\bibitem[\protect\citeauthoryear{Liu \bgroup \em et al.\egroup
  }{2023}]{liu2023resdo}
Yanhong Liu, Ji~Shen, Lei Yang, Guibin Bian, and Hongnian Yu.
\newblock Resdo-unet: A deep residual network for accurate retinal vessel
  segmentation from fundus images.
\newblock {\em Biomedical Signal Processing and Control}, 79:104087, 2023.

\bibitem[\protect\citeauthoryear{Long \bgroup \em et al.\egroup
  }{2015}]{long2015fully}
Jonathan Long, Evan Shelhamer, and Trevor Darrell.
\newblock Fully convolutional networks for semantic segmentation.
\newblock In {\em Proceedings of the IEEE conference on computer vision and
  pattern recognition}, pages 3431--3440, 2015.

\bibitem[\protect\citeauthoryear{Lv \bgroup \em et al.\egroup }{2022}]{lv20222}
Peiqing Lv, Jinke Wang, and Haiying Wang.
\newblock 2.5 d lightweight riu-net for automatic liver and tumor segmentation
  from ct.
\newblock {\em Biomedical Signal Processing and Control}, 75:103567, 2022.

\bibitem[\protect\citeauthoryear{Milletari \bgroup \em et al.\egroup
  }{2016}]{milletari2016v}
Fausto Milletari, Nassir Navab, and Seyed-Ahmad Ahmadi.
\newblock V-net: Fully convolutional neural networks for volumetric medical
  image segmentation.
\newblock In {\em 2016 fourth international conference on 3D vision (3DV)},
  pages 565--571. Ieee, 2016.

\bibitem[\protect\citeauthoryear{Oktay \bgroup \em et al.\egroup
  }{2018}]{oktay2018attention}
Ozan Oktay, Jo~Schlemper, Loic~Le Folgoc, Matthew Lee, Mattias Heinrich,
  Kazunari Misawa, Kensaku Mori, Steven McDonagh, Nils~Y Hammerla, Bernhard
  Kainz, et~al.
\newblock Attention u-net: Learning where to look for the pancreas.
\newblock {\em arXiv preprint arXiv:1804.03999}, 2018.

\bibitem[\protect\citeauthoryear{Ronneberger \bgroup \em et al.\egroup
  }{2015}]{ronneberger2015u}
Olaf Ronneberger, Philipp Fischer, and Thomas Brox.
\newblock U-net: Convolutional networks for biomedical image segmentation.
\newblock In {\em Medical Image Computing and Computer-Assisted
  Intervention--MICCAI 2015: 18th International Conference, Munich, Germany,
  October 5-9, 2015, Proceedings, Part III 18}, pages 234--241. Springer, 2015.

\bibitem[\protect\citeauthoryear{Shamshad \bgroup \em et al.\egroup
  }{2023}]{shamshad2023transformers}
Fahad Shamshad, Salman Khan, Syed~Waqas Zamir, Muhammad~Haris Khan, Munawar
  Hayat, Fahad~Shahbaz Khan, and Huazhu Fu.
\newblock Transformers in medical imaging: A survey.
\newblock {\em Medical Image Analysis}, page 102802, 2023.

\bibitem[\protect\citeauthoryear{Song \bgroup \em et al.\egroup
  }{2020}]{song2020real}
Jiarui Song, Beibei Li, Yuhao Wu, Yaxin Shi, and Aohan Li.
\newblock Real: a new resnet-alstm based intrusion detection system for the
  internet of energy.
\newblock In {\em 2020 IEEE 45th Conference on Local Computer Networks (LCN)},
  pages 491--496. IEEE, 2020.

\bibitem[\protect\citeauthoryear{Suetens}{2017}]{suetens2017fundamentals}
Paul Suetens.
\newblock {\em Fundamentals of medical imaging}.
\newblock Cambridge university press, 2017.

\bibitem[\protect\citeauthoryear{Taha and Hanbury}{2015}]{taha2015metrics}
Abdel~Aziz Taha and Allan Hanbury.
\newblock Metrics for evaluating 3d medical image segmentation: analysis,
  selection, and tool.
\newblock {\em BMC medical imaging}, 15(1):1--28, 2015.

\bibitem[\protect\citeauthoryear{Tang \bgroup \em et al.\egroup
  }{2022}]{tang2022self}
Yucheng Tang, Dong Yang, Wenqi Li, Holger~R Roth, Bennett Landman, Daguang Xu,
  Vishwesh Nath, and Ali Hatamizadeh.
\newblock Self-supervised pre-training of swin transformers for 3d medical
  image analysis.
\newblock In {\em Proceedings of the IEEE/CVF Conference on Computer Vision and
  Pattern Recognition}, pages 20730--20740, 2022.

\bibitem[\protect\citeauthoryear{Valanarasu \bgroup \em et al.\egroup
  }{2021}]{valanarasu2021medical}
Jeya Maria~Jose Valanarasu, Poojan Oza, Ilker Hacihaliloglu, and Vishal~M
  Patel.
\newblock Medical transformer: Gated axial-attention for medical image
  segmentation.
\newblock In {\em Medical Image Computing and Computer Assisted
  Intervention--MICCAI 2021: 24th International Conference, Strasbourg, France,
  September 27--October 1, 2021, Proceedings, Part I 24}, pages 36--46.
  Springer, 2021.

\bibitem[\protect\citeauthoryear{Vaswani \bgroup \em et al.\egroup
  }{2017}]{vaswani2017attention}
Ashish Vaswani, Noam Shazeer, Niki Parmar, Jakob Uszkoreit, Llion Jones,
  Aidan~N Gomez, {\L}ukasz Kaiser, and Illia Polosukhin.
\newblock Attention is all you need.
\newblock {\em Advances in neural information processing systems}, 30, 2017.

\bibitem[\protect\citeauthoryear{Wang \bgroup \em et al.\egroup
  }{2021a}]{wang2021boundary}
Jiacheng Wang, Lan Wei, Liansheng Wang, Qichao Zhou, Lei Zhu, and Jing Qin.
\newblock Boundary-aware transformers for skin lesion segmentation.
\newblock In {\em Medical Image Computing and Computer Assisted
  Intervention--MICCAI 2021: 24th International Conference, Strasbourg, France,
  September 27--October 1, 2021, Proceedings, Part I 24}, pages 206--216.
  Springer, 2021.

\bibitem[\protect\citeauthoryear{Wang \bgroup \em et al.\egroup
  }{2021b}]{wang2021pyramid}
Wenhai Wang, Enze Xie, Xiang Li, Deng-Ping Fan, Kaitao Song, Ding Liang, Tong
  Lu, Ping Luo, and Ling Shao.
\newblock Pyramid vision transformer: A versatile backbone for dense prediction
  without convolutions.
\newblock In {\em Proceedings of the IEEE/CVF international conference on
  computer vision}, pages 568--578, 2021.

\bibitem[\protect\citeauthoryear{Wang \bgroup \em et al.\egroup
  }{2021c}]{wang2021crossformer}
Wenxiao Wang, Lu~Yao, Long Chen, Binbin Lin, Deng Cai, Xiaofei He, and Wei Liu.
\newblock Crossformer: A versatile vision transformer hinging on cross-scale
  attention.
\newblock {\em arXiv preprint arXiv:2108.00154}, 2021.

\bibitem[\protect\citeauthoryear{Wang \bgroup \em et al.\egroup
  }{2022}]{wang2022uctransnet}
Haonan Wang, Peng Cao, Jiaqi Wang, and Osmar~R Zaiane.
\newblock Uctransnet: rethinking the skip connections in u-net from a
  channel-wise perspective with transformer.
\newblock In {\em Proceedings of the AAAI conference on artificial
  intelligence}, volume~36, pages 2441--2449, 2022.

\bibitem[\protect\citeauthoryear{Wu \bgroup \em et al.\egroup
  }{2021}]{wu2021cvt}
Haiping Wu, Bin Xiao, Noel Codella, Mengchen Liu, Xiyang Dai, Lu~Yuan, and Lei
  Zhang.
\newblock Cvt: Introducing convolutions to vision transformers.
\newblock In {\em Proceedings of the IEEE/CVF International Conference on
  Computer Vision}, pages 22--31, 2021.

\bibitem[\protect\citeauthoryear{Yang \bgroup \em et al.\egroup
  }{2022}]{yang2022dcu}
Xin Yang, Zhiqiang Li, Yingqing Guo, and Dake Zhou.
\newblock Dcu-net: a deformable convolutional neural network based on cascade
  u-net for retinal vessel segmentation.
\newblock {\em Multimedia Tools and Applications}, 81(11):15593--15607, 2022.

\bibitem[\protect\citeauthoryear{Zhou \bgroup \em et al.\egroup
  }{2018}]{zhou2018unet++}
Zongwei Zhou, Md~Mahfuzur Rahman~Siddiquee, Nima Tajbakhsh, and Jianming Liang.
\newblock Unet++: A nested u-net architecture for medical image segmentation.
\newblock In {\em Deep Learning in Medical Image Analysis and Multimodal
  Learning for Clinical Decision Support: 4th International Workshop, DLMIA
  2018, and 8th International Workshop, ML-CDS 2018, Held in Conjunction with
  MICCAI 2018, Granada, Spain, September 20, 2018, Proceedings 4}, pages 3--11.
  Springer, 2018.

\end{thebibliography}

\end{document}